\documentclass[preprint,12pt]{elsarticle}
\usepackage{graphicx}
\usepackage{epsfig}
\usepackage{amssymb,amsmath}
\usepackage{multicol}
\RequirePackage[dvipsnames,usenames]{color}
\journal{Physics Letters B}
\newcommand{\mpl}{m_{\rm P}}
\begin{document}
\begin{frontmatter}
\title{\textbf{Hawking radiation and the Bloom-Gilman duality}}
\author{R.~Casadio$^{1}$}\ead{Roberto.Casadio@bo.infn.it}
\author{A.Yu.~Kamenshchik$^{1,2}$}\ead{Alexander.Kamenshchik@bo.infn.it}
\author{O.V.~Teryaev$^{3,4}$}\ead{teryaev@theor.jinr.ru}  
\address{
\small $^1$Dipartimento di Fisica e Astronomia, Universit\`a di Bologna, and INFN,
\\
Via Irnerio~46, 40126~Bologna, Italy.
\\
\small $^2$L.D.~Landau Institute for Theoretical Physics of the Russian Academy of Sciences,
\\
Kosygin str.~2, 119334~Moscow, Russia.
\\
\small $^3$Joint Institute for Nuclear Research, 
141980 Dubna, Russia.
\\
\small $^4$ Lomonosov Moscow State University,
\\
Leninskie Gory~1, 119991~Moscow, Russia}
%
\date{ \ }
\begin{abstract}
The decay widths of the quantum black hole precursors, determined from the poles
of the resummed graviton propagator, are matched to the expected lifetime 
given by the Hawking decay.
In this way, we impose a sort of duality between a perturbative description and an
essentially non-perturbative description, bearing some similarity with the Bloom-Gilman
duality for the strong interactions.   
General relations are then obtained for the widths and masses of the poles in
terms of the number of particle species and the renormalisation scale of gravity.
\end{abstract}
\end{frontmatter}
\section{Introduction}
Black holes are non-perturbative objects that arise in General Relativity, and describe a strong
regime of gravity in which all signals are classically confined within their horizon.
Like for other bound states, one can hope to find hints of their existence already at the perturbative
level of quantum field theory.
In fact, the resummed one-loop propagator of the graviton interacting with matter fields obtained in
Ref.~\cite{grav-prop,grav-prop1} contains non-trivial poles, which may be interpreted as precursors of
black holes~\cite{Calmet,Casadio}.
\par
The complete analytic structure of this propagator was further studied in Ref.~\cite{we}, where a broad
spectrum of such resonance-like states was found.
In particular, the resummed graviton propagator has the following form:
\begin{equation}
i\,D^{\alpha\beta}(p^2)
=
i
\left(L^{\alpha\mu}L^{\beta\nu}+L^{\alpha\nu}L^{\beta\mu}-L^{\alpha\beta}L^{\mu\nu}\right)
G(p^2)
\ ,
\label{prop}
\end{equation}
where 
\begin{equation}
L^{\mu\nu}(p)
=
\eta^{\mu\nu}-\frac{p^{\mu}p^{\nu}}{p^2}
\label{prop1}
\end{equation}
and 
\begin{equation}
G^{-1}(p^2)
=
2\,p^2
\left[1-\frac{N\,p^2}{120\,\pi\, \mpl^2}\ln\!\left(-\frac{p^2}{\mu^2}\right)\right]
\ .
\label{prop2}
\end{equation}
Here, $\mpl$ denotes the Planck mass, $\mu$ is the renormalization scale,
$N=N_s+3\,N_f+12\,N_V$, where $N_s$, $N_f$, $N_V$
are the number of scalar, fermion and vector fields, respectively.
In the Standard Model of particle physics, $N_s=4$, $N_f = 45$, $N_V=12$ and $N=283$.
The propagator~(\ref{prop}) has a standard pole at $p^2=0$ and an infinite
number of other poles, which are the zeros of the expression~(\ref{prop2}).
The masses of these resonance-like states are given by~\cite{we}
\begin{equation}
m_n
=
\mpl\,
\sqrt{\frac{120\,\pi}{N}\,\frac{\sin\theta_n}{\theta_n}}
\left|\sin\!\left(\frac{\theta_n}{2}\right)\right|
\ ,
\label{mGam6}
\end{equation}
and the corresponding widths by
\begin{equation}
\Gamma_n
=
\mpl\,\sqrt{\frac{120\,\pi}{N}\,\frac{\sin\theta_n}{\theta_n}}\,
\frac{\sin\theta_n}{\left|\sin({\theta_n}/{2})\right|}
\ ,
\label{mGam7}
\end{equation}
where the integer $n\ge 0$ labels the $n^{\rm th}$ Riemann sheet.
Their ratio is thus 
\begin{equation}
\frac{\Gamma_n}{m_n}
=
2\,\cot\!\left(\frac{\theta_n}{2}\right)
\ ,
\label{GsM}
\end{equation}
where the angle $\theta_n$ is the solution of the equation
\begin{equation}
\frac{\theta_n}{\sin\theta_n}\,
\exp\!\left(-\frac{\theta_n}{\tan\theta_n}\right)
=
\frac{120\, \pi\, \mpl^2}{N\, \mu^2}
\ ,
\label{theta}
\end{equation}
defining the phase of the resonance on the $n^{\rm th}$ Riemann sheet.
It is important to recall that in Ref.~\cite{we} we showed that the phase $\theta$ should belong
to one of the intervals
\begin{equation}
2\,\pi\, n \leq \theta_n \leq (2\,n +1)\,\pi
\ ,
\label{int}
\end{equation}
for these solutions to represent proper resonances.
\par
One may nevertheless ask why such an essentially non-perturbative object
as a quantum black hole can be described by means of the perturbation theory.
This problem inevitably arises in quantum chromodynamics (QCD), where, due to the
confinement property, the observed objects are non-perturbatively formed hadrons,
whereas the actual calculations can be performed at the level of quarks and gluons.
The central role here is played by various types of quark-hadron duality.
The first one was probably the famous Bloom-Gilman duality~\cite{BG-duality} between 
parton distributions and hadronic resonances, which was marked by Feynman as a
manifestation of Bohr's complementarity~\cite{Feynman}.    
The quark-hadron duality is crucial in the applications of QCD sum rules~\cite{Shifman},
where the non-perturbative vector-meson coupling can be related to the simple quark loop.
The quark-hadron duality is also related to such a fundamental quantity as the axial
anomaly~\cite{Teryaev}.
Recently, a similar duality was observed in relativistic hydrodynamics
as an effective theory~\cite{Teryaev1}.  
\par   
In the present paper, we conjecture that a quantum black hole is related to perturbative
matter loops like, for example, the $\rho$-meson is related to quark loops. 
To do so, we confront our previous results~\cite{we}, based on perturbative
calculations~\cite{grav-prop,grav-prop1}, with the non-perturbative description  
of the Hawking evaporation~\cite{Hawking}.
Namely, we postulate that the widths~\eqref{mGam7} determined by the poles of the
graviton propagator~\eqref{prop} equal the inverse of the decay times following from
the canonical Hawking evaporation.
This assumption leads to general relations between the parameters that describe
the black hole precursors.
\section{Duality for black holes}
For a quantum black hole, that is a black hole with mass of the order of the Planck mass,
it should be possible to estimate the decay width from its finite lifetime $\tau$ due to Hawking
radiation~\cite{Hawking}, that is
\begin{equation}
\Gamma_{\rm H}
= \tau^{-1}=
\frac{\mpl^4}{\alpha \, m^3}
\ ,
\label{GamH}
\end{equation}
where $\alpha$ is a positive dimensionless parameter, which depends on the unknown details
of the Hawking emission from such extreme black holes.
For example, in the case of the standard thermodynamic approach exploiting the Stefan-Boltzmann
law (see {\em e.g.}~ Ref.~\cite{Walecka:2008zz,FL})
this coefficient is rather large, that is
\begin{equation}
\alpha_{\rm SB}
=
5120\, \pi
\ .
\label{aSB}
\end{equation}
Now, we make the strong assumption that this width should coincide with the one
given in Eq.~\eqref{mGam7},
\begin{equation}
\Gamma_{\rm H}
=
\Gamma_n
\ .
\label{axiom}
\end{equation}
Substituting~\eqref{mGam7} and \eqref{GamH} into \eqref{axiom} yields
\begin{equation}
F(\theta)
\equiv \frac{\sin^3(\theta)\,\sin^2(\theta/2)}{\theta^2}
=
A
\ ,
\label{dual}
\end{equation}
with
\begin{equation}
A
\equiv
\frac{N^2}{(120\,\pi)^2\,\alpha}
\simeq
7.0\cdot 10^{-6}\,\frac{N^2}{\alpha}
\ ,
\end{equation}
which gives $A\simeq 0.56/\alpha^2$ for $N=283$.
This is the value of $N$ we will mostly refer to, but we already notice that $A$ is very
sensitive to the actual particle content of the theory.
\begin{figure}[t]
\centerline{\epsfxsize 10cm
\epsfbox{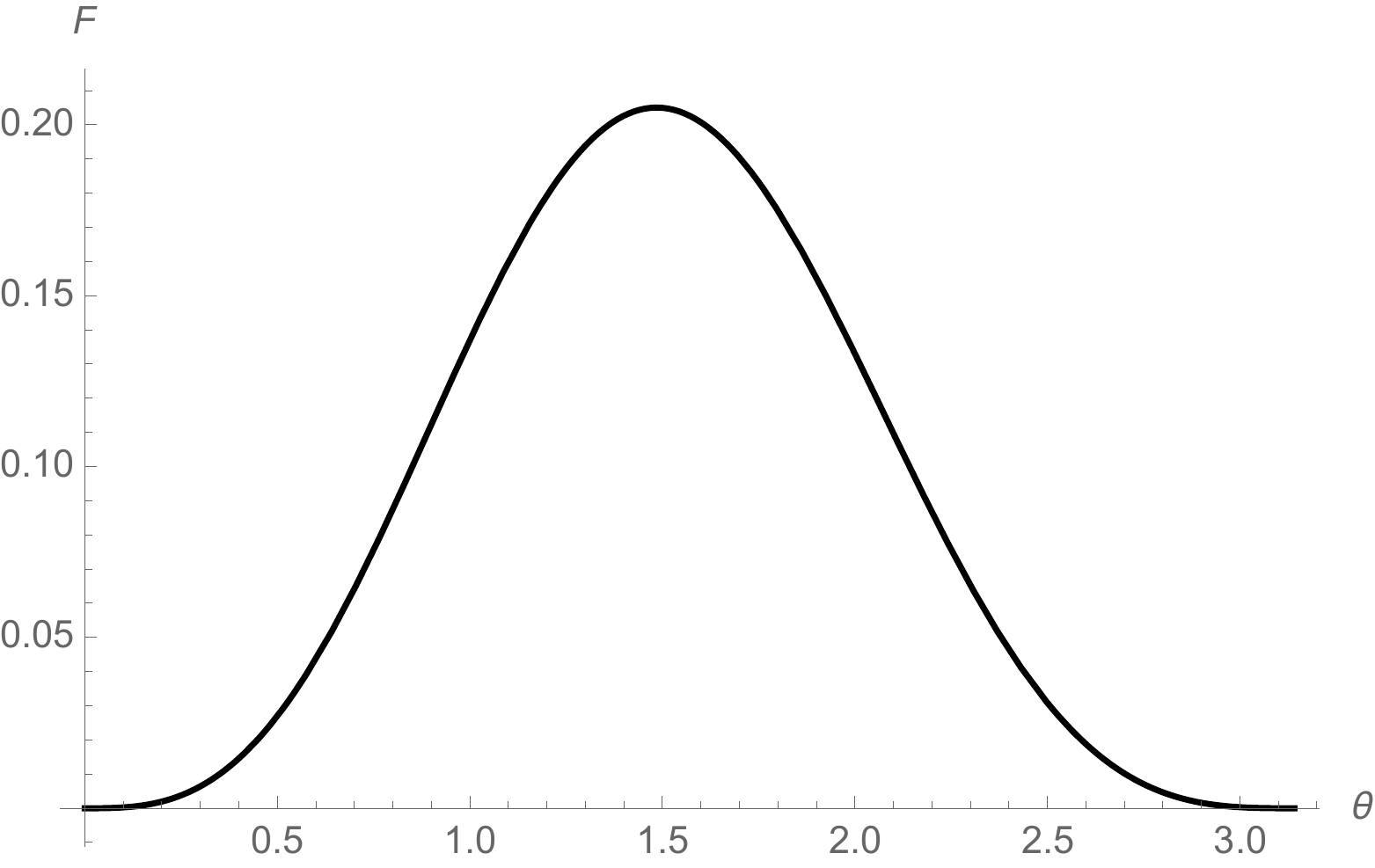}}
\caption{The function $F(\theta)$ in the allowed interval $0\le\theta\le\pi$
on the first sheet of the Riemann surface.
\label{F0}}
\end{figure}
\par
The first interesting thing to note is that there exists a minimum
value of $\alpha$ for which Eq.~\eqref{dual} admits a solution given a fixed number
$N$ of matter particles.
In fact, the function $F$ inside the allowed intervals~\eqref{int} 
is a bell-shaped curve like the one shown in Fig.~\ref{F0} for the Riemann sheet $n=0$.
For larger values of $n$, the shape is similar, except the maximum value
of $F$ simply decreases like $\theta^{-2}\sim n^{-2}$.
The absolute maximum which occurs for $\theta$ in the $n=0$ sheet
can be easily found to be $F_{\rm max}\equiv F(\theta_{\rm max}\simeq 1.48)\simeq 0.21$.
This implies that one must have $A\lesssim F_{\rm max}$ or
\begin{equation}  
\alpha
\gtrsim
4.9\,\frac{(120\,\pi)^2}{N^2}
\simeq
2.7
\ ,
\label{a_min}
\end{equation}
for a solution $\theta_0$ to exist.
For $A<F_{\rm max}$, one will have two such solutions, say $\theta_0^\pm$,
which degenerate to one for $A= F_{\rm max}$, that is $\theta_0^\pm=\theta_{\rm max}$.
This case corresponds to 
\begin{equation}
m_0
\simeq
0.64\,\mpl
\  ,
\qquad
\Gamma_0
\simeq
1.4\,\mpl
\ , 
\end{equation}
and, from Eq.~\eqref{theta}, a renormalisation scale $\mu_0\simeq 0.3\,\mpl$.
\begin{figure}[t]
\centerline{\epsfxsize 10cm
\epsfbox{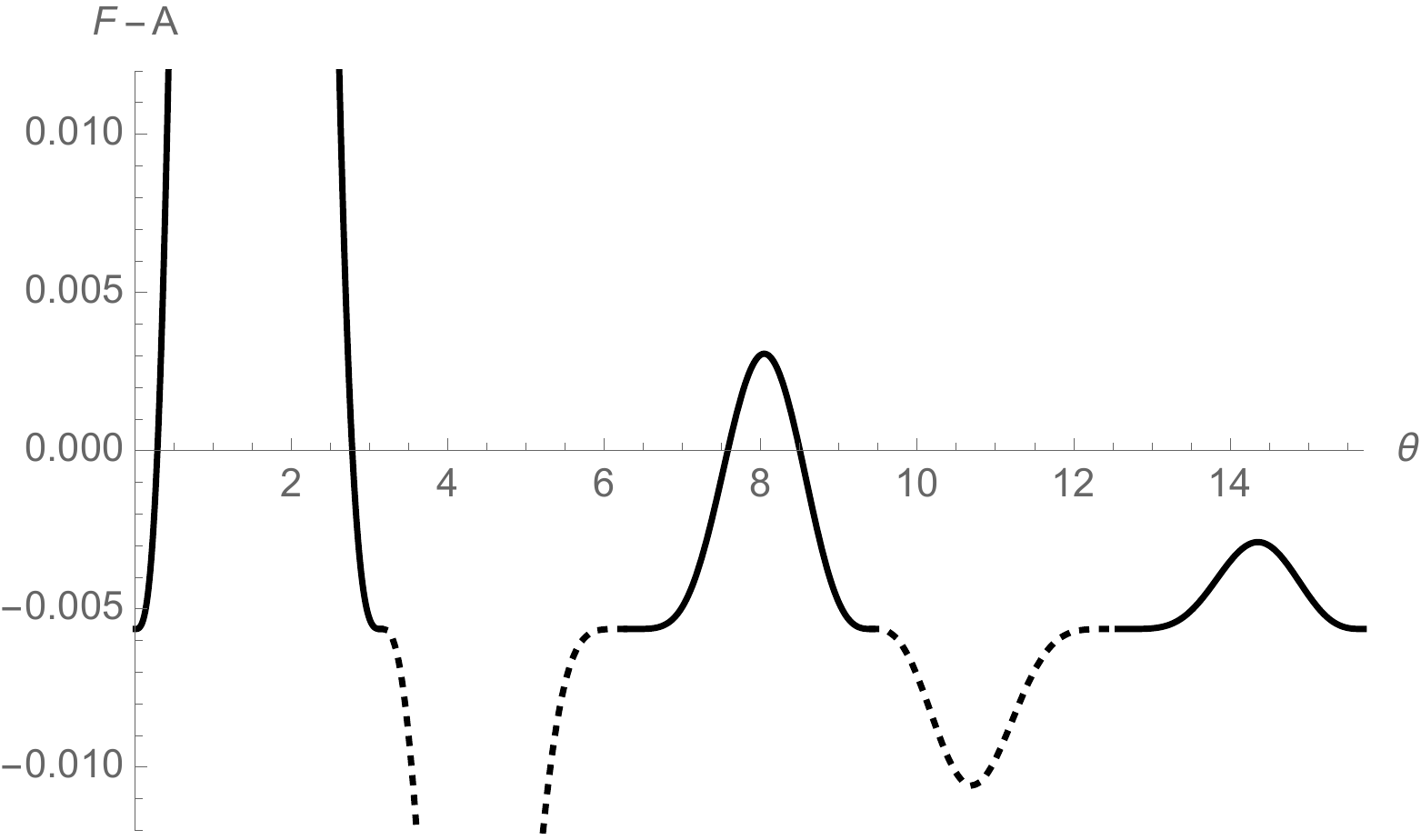}}
\caption{The function $F(\theta)-A$ for $\alpha=100$, in the allowed intervals $2\,n\,\pi\le\theta\le (2\,n+1)\,\pi$
(solid line) for $n=0,1,2$. 
Phases $\theta_n^\pm$ are given by intersections with the $\theta$ axis.
Dotted line is the same function in the forbidden intervals.
\label{F100}}
\end{figure}
\par
\begin{figure}[t]
\centerline{\epsfxsize 10cm
\epsfbox{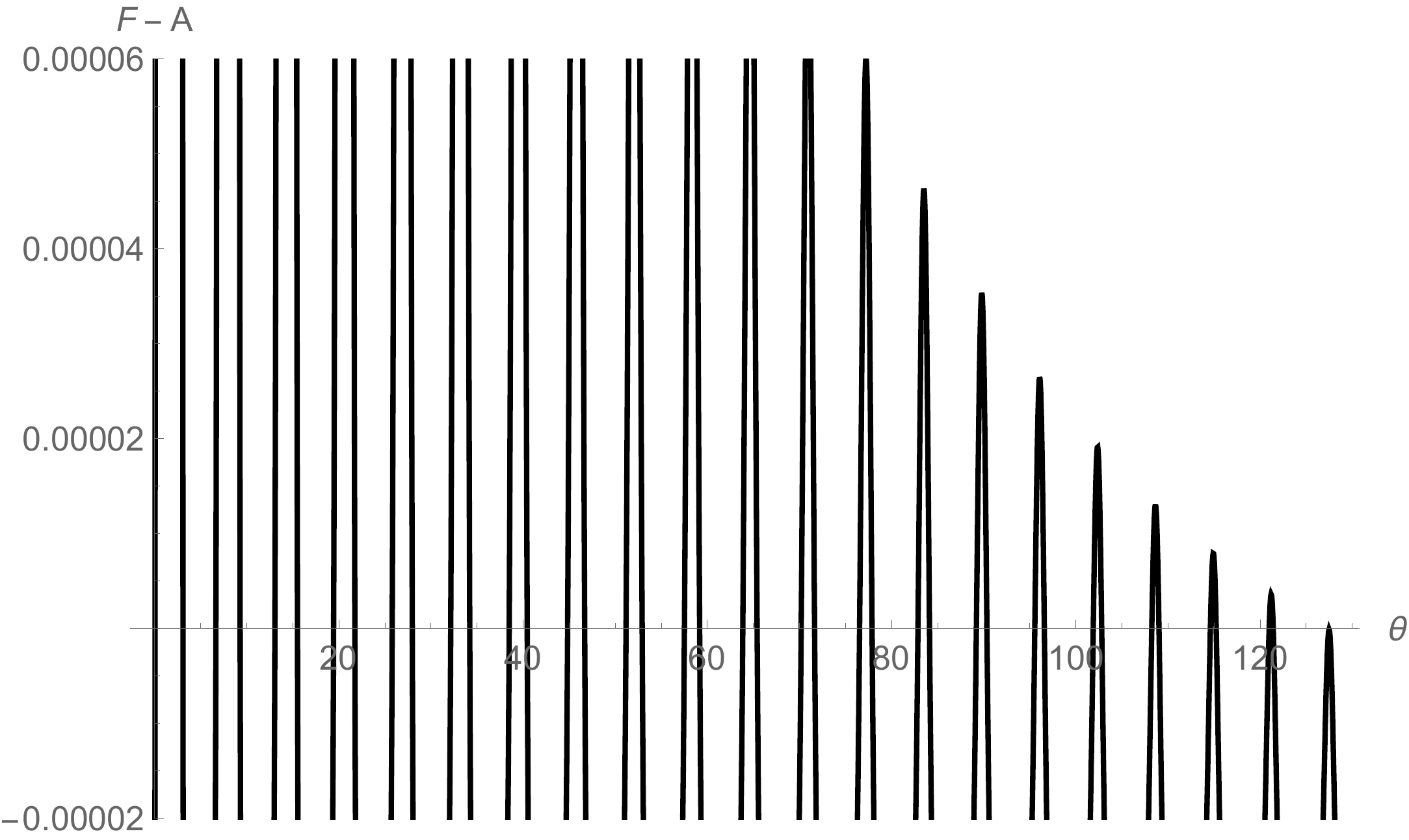}}
\caption{The function $F(\theta)-A$ for $\alpha=\alpha_{\rm SB}$, in the interval $0\le\theta\le 41\,\pi$
(or $n=0,1,\ldots,20$). 
Phases $\theta_n^\pm$ are given by intersections with the $\theta$ axis and exist up to $n=19$ included.
\label{FSB}}
\end{figure}

For sufficiently small $A$ (equivalently, for sufficiently large $\alpha$),
one can find solutions in Riemann sheets with $n>0$, and the number of sheets with solutions
increases for increasing $\alpha$.
For instance, for $\alpha=100$ one has two solutions $\theta_0^\pm$ in the first Riemann sheet,
and two more solutions $\theta_1^\pm$ in the second Riemann sheet, as can be seen from
Fig.~\ref{F100}.
The corresponding relevant physical quantities are displayed in Table~\ref{T100}.
Likewise, for $\alpha=\alpha_{\rm SB}$ in Eq.~\eqref{aSB}, one can see from Fig.~\ref{FSB}
that solutions $\theta_n^\pm$ exist up to $n=19$, and we display a few in Table~\ref{TSB}.
\begin{table}[h]
\centering
\begin{tabular}{|c|c|c|c|c|}
\hline
$\alpha=100$ & $m$ & $\Gamma$ & $\Gamma/m$ & $\mu$
 \\
 \hline
 \hline
 $\theta_0^-=0.29$ & 0.16 & 2.3 & 14 & 0.6
 \\
 \hline
 $\theta_0^+=2.8$ & 0.41 & 0.15 & 0.4 & $3\cdot 10^{-3}$
 \\
 \hline
 $\theta_1^-=7.6$ & 0.25 & 0.66 & 2.6 & 0.4
 \\
 \hline
 $\theta_1^+=8.5$ & 0.32 & 0.32 & 1 & $5\cdot 10^{-3}$
 \\
 \hline
\end{tabular}
\caption{Phases, and corresponding masses and widths (in units of $\mpl$), for $\alpha=100$.}
\label{T100}
\end{table}
\begin{table}[h]
\centering
\begin{tabular}{|c|c|c|c|c|}
\hline
$\alpha=\alpha_{\rm SB}$ & $m$ & $\Gamma$ & $\Gamma/m$ & $\mu$
 \\
 \hline
 \hline
 $\theta_0^-=5.2\cdot 10^{-2}$ & $3.0\cdot 10^{-2}$ & 2.3 & 77 & 0.6
 \\
 \hline
 $\theta_0^+=3.1$ & $1.7\cdot 10^{-1}$ & $1.5\cdot 10^{-1}$ & $7.0\cdot 10^{-2}$ & $1\cdot 10^{-11}$
  \\
 \hline
  $\theta_{19}^-=121.0$ & $7.5\cdot 10^{-2}$ & $1.5\cdot 10^{-1}$ & 1.9 & $7\cdot 10^{-3}$
 \\
 \hline
 $\theta_{19}^+=121.4$ & $8.4\cdot 10^{-2}$ &$1.0\cdot 10^{-1}$ & 1.2 & $2\cdot 10^{-15}$
 \\
 \hline
\end{tabular}
\caption{Phases for lowest and highest $n$, and corresponding masses and widths
(in units of $\mpl$), for $\alpha=\alpha_{\rm SB}$.}
\label{TSB}
\end{table}
\par
We can now present some general considerations.
It is easy to see from Eq.~\eqref{GsM} that the ratio $\Gamma_n/m_n$ is minimised  
for $\theta_n^+\simeq (2\,n+1)\,\pi$ inside the allowed intervals~\eqref{int}.
For any given value of $A$ that allows for the existence of resonances, $\theta_0^+$ is always
the largest phase (modulo $2\,\pi$), and will therefore correspond to the (relatively) most stable resonance.
On the contrary, since $\theta_0^-$ is the closest to $0$ (modulo $2\,\pi$), the corresponding resonance
will always be the (relatively) most unstable.
Moreover, the larger $\alpha$ (and thus the larger the number of sheets with solutions), the larger is
$\theta_0^+$ (modulo $2\,\pi$), which makes it more stable  (conversely, $\theta_0^-$ is more unstable).
We also notice that $\Gamma_n/m_n\simeq 2$ for $\theta_n\simeq \pi/2$ (modulo $2\,\pi$),
which can be a solution only provided $A$ is close to the maximum of $F$.
This will occur if $n$ is the largest integer that admits solutions, so that $\theta_0^-\simeq \theta_n^+$.
Such properties are clearly displayed by the cases we considered explicitly above.
\section{Conclusions}
In this work, we have explored the opportunity to apply ideas similar to the well-known
quark-hadron duality to the poles of the resummed graviton propagator~\eqref{prop},
which are interpreted as describing the smallest possible black holes.
\par
There is some analogy between the perturbative and non-perturbative physics 
in QCD and gravity.
The contribution of quark loops to the photon propagator~\cite{Shifman} allows one
to describe the properties of vector mesons in QCD.
At the same time, the contribution of matter loops to the graviton propagator is related to
properties of black holes.
\par
Note that perturbative QCD is not sufficient to describe the form of the spectral function.
Indeed, in order to establish the duality quantitatively and its domain of validity,
the notion of vacuum condensates is also required~\cite{Shifman}. 
At the same time, the perturbative loop contributions are already sufficient to identify
the structures which may be interpreted as Breit-Wigner peaks. 
\par
The origin of duality is connected with unitarity and the possibility to have different choices
for a full set of states.
In QCD, for example, one can use the fundamental (quark and gluon) states or
the physical (hadron) states as such full sets, since they are complementary to each other
and {\it not additive\/} (for the notions of duality and additivity for various types of QCD
factorisation, see Ref.~\cite{Anikin:2008bq}).
The duality for black holes would mean that the matter states and the states of quantum
black holes are also complementary and one should not need to consider both of them
together. 
\par
The emerging analytical structure of the graviton propagator provides a rich set of
black hole states.
It is possible that they indeed form a full set of states and provide a complementary
view to the quantum theory.
\section*{Acknowledgments.}
O.T.~thanks the Department of Physics and Astronomy and INFN section of Bologna
for the kind hospitality during the development of this work.
R.C.~and A.K.~are partially supported by the INFN grant FLAG.
A.K.~was partially supported by the RFBR grant N.~17-02-01008. 
O.T.~was partially supported by the RFBR grant N.~14-01-00647.
The work of R.C.~has been carried out in the framework 
of GNFM and INdAM and the COST action {\em Cantata\/}. 
\end{document}